\documentclass[pre,twocolumn,showpacs,amsmath,amssymb]{revtex4}
\usepackage{epsfig} 

\newcommand{\bsub}{\begin{subequations}}
\newcommand{\esub}{\end{subequations}}
\newcommand{\calA}{\mathcal A}
\newcommand{\calC}{\mathcal C}
\newcommand{\calP}{\mathcal P}

\newcommand{\ez}{\mathbf{e^{{}}}_z}
\newcommand{\vvv}{\mathbf{v^{{}}}}
\newcommand{\vz}{v}
\newcommand{\pp}{\partial^{{}}}
\newcommand{\ppsqr}{\partial^{2_{}}}
\newcommand{\Rhyd}{R^{{}}_\textrm{hyd}}
\newcommand{\RhydS}{R^{*_{}}_\textrm{hyd}}

\begin{document}

\title{Reexamination of Hagen--Poiseuille flow:\\ shape-dependence of the hydraulic
resistance in microchannels}

\author{Niels Asger Mortensen, Fridolin Okkels, and Henrik Bruus}

\affiliation{MIC -- Department of Micro and Nanotechnology,
bldg.~345~east\\ Technical University of Denmark, DK-2800
Kgs.~Lyngby, Denmark.}

\date{February 3, 2004}

\begin{abstract}
We consider pressure-driven, steady state Poiseuille flow in
straight channels with various cross-sectional shapes: elliptic,
rectangular, triangular, and harmonic-perturbed circles. A given
shape is characterized by its perimeter $\calP$ and area $\calA$
which are combined into the dimensionless compactness number
$\calC = \calP^2/\calA$, while the hydraulic resistance is
characterized by the well-known dimensionless geometrical
correction factor $\alpha$. We find that $\alpha$ depends linearly
on $\calC$, which points out $\calC$ as a single dimensionless
measure characterizing flow properties as well as the strength and
effectiveness of surface-related phenomena central to
lab-on-a-chip applications. This measure also provides a simple
way to evaluate the hydraulic resistance for the various shapes.
\end{abstract}

 \maketitle

\section{Introduction}
\label{sec:introduction}

\begin{figure}[b!]
\begin{center}
\epsfig{file=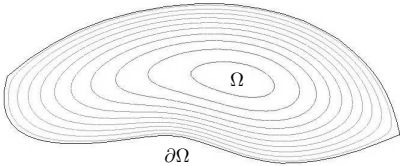, width=0.3\textwidth,clip}
\end{center}
\caption{\label{fig:geometry} An arbitrary cross-sectional shape
$\Omega$ with perimeter $\pp\Omega$ of a straight fluid channel
with pressure-driven steady-state flow. The contours show the
velocity $\vz(x,y)$ obtained numerically from
Eq.~(\ref{eq:Poisson}) by a finite-element method. The velocity is
zero at the boundary and maximal near the centre-of-mass. }
\end{figure}

The rapid development in the field of lab-on-a-chip systems during
the past decade has put emphasis on studies of shape-dependence in
microfluidic channels. Traditionally, capillary tubes would have
circular cross-sections, but today microfabricated channels have a
variety of shapes depending on the fabrication technique in use.
Examples are rectangular channels obtained by hot embossing in
polymer wafers, semi-circular channels in isotropically etched
surfaces, triangular channels in KOH-etched silicon crystals,
Gaussian-shaped channels in laser-ablated polymer films, and
elliptic channels in stretched PDMS devices, see e.g.,
Ref.~\onlinecite{Geschke:04a}.

The pressure-driven, steady-state flow of a liquid through long,
straight, and rigid channels of any constant cross-sectional shape
is referred to as Hagen--Poiseuille (or simply Poiseuille) flow,
and it is often characterized by the hydraulic resistance, $\Rhyd
= \Delta p/Q$, where $\Delta p$ is the pressure drop along the
channel and $Q$ the flow rate through the channel. In
Fig.~\ref{fig:geometry} is shown an arbitrarily shaped
cross-section $\Omega$ in the $xy$ plane for a straight channel
placed along the $z$ axis. A natural unit for the hydraulic
resistance is given by dimensional analysis as $\RhydS \equiv \eta
L/\calA^2$, where $L$ is the channel length, $\eta$ the dynamic
viscosity of the liquid, and $\calA = \int_\Omega dxdy$ the
cross-sectional area. Typically, the fluid flow is subject to a
no-slip boundary condition at the walls $\pp\Omega$ and thus the
actual hydraulic resistance will depend on the perimeter as well
as the cross-section area. This dependence can therefore be
characterized by the dimensionless geometrical correction factor
$\alpha$ given by
 \begin{equation} \label{eq:alphaDef}
 \alpha \equiv \frac{\Rhyd}{\RhydS}.
 \end{equation}
In lab-on-a-chip applications~\cite{Geschke:04a,Sanders:00a},
where large surface-to-volume ratios are encountered, the problem
of the bulk Poiseuille flow is typically accompanied by other
surface-related physical or bio-chemical phenomena in the fluid.
The list of examples includes surface chemistry, DNA hybridization
on fixed targets, catalysis, interfacial electrokinetic phenomena
such as electro-osmosis, electrophoresis and electro-viscous
effects as well as continuous edge-source diffusion. Though the
phenomena are of very different nature, they have at least one
thing in common; they are all to some degree surface phenomena and
their strength and effectiveness depends strongly on the
surface-to-volume ratio. It is common to quantify this by the
dimensionless compactness $\calC$ given by
 \begin{equation} \label{eq:CDef}
 \calC \equiv \frac{\calP^2}{\calA},
 \end{equation}
where $\calP \equiv \int_{\partial\Omega}d\ell$ is the perimeter
of the boundary $\partial \Omega$ confining the fluid, see
Fig.~\ref{fig:geometry}. For other measures of $\calC$ we refer to
Ref.~\onlinecite{Bogaert:00a} and references therein. In this
paper we demonstrate a simple dependence of the geometrical
correction factor $\alpha$ on the compactness $\calC$ and our
results thus point out a unified dimensionless measure of flow
properties as well as the strength and effectiveness of
surface-related phenomena central to lab-on-a-chip applications.
Furthermore, our results allow for an easy evaluation of the
hydraulic resistance for elliptical, rectangular, and triangular
cross-sections with the geometrical measure $\calC$ being the only
input parameter. Above we have emphasized microfluidic flows
because here a variety of shapes are frequently encountered.
However, our results are generally valid for all laminar flows.

\section{Poiseuille flow}
Due to translation invariance along the $z$ axis the velocity
field of a Newtonian fluid in a straight channel is parallel to
the $z$ axis, and takes the form $\vvv = \vz(x,y)\ez$.
Consequently, the non-linear term in the Navier--Stokes equation
drops out~\cite{Landau:87a}, and in steady-state, given the
pressure gradient $-(\Delta p/L)\ez$, the velocity $\vz(x,y)$ is
thus given by the Poisson equation,
\begin{equation}\label{eq:Poisson}
 \big(\ppsqr_x + \ppsqr_y\big)\vz(x,y)
 = \frac{\Delta p}{\eta L},
 \end{equation}
with the velocity being subject to a no-slip condition at the
boundary $\partial\Omega$.
The relation between the pressure drop $\Delta p$, the velocity
$\vz(x,y)$, and the geometrical correction factor $\alpha$ becomes
 \begin{equation}\label{eq:Deltap}
 \Delta p = \Rhyd Q = \alpha \RhydS Q =
 \alpha \RhydS \int_\Omega dxdy\,\vz(x,y),
 \end{equation}
where $Q$ is the volume flow rate.

\section{The geometrical correction factor versus compactness}
Our main objective is to find the relation between the geometrical
correction factor $\alpha$ and the compactness $\calC$ for various
families of geometries.

\subsection{Elliptical cross section}
\begin{figure}[t!]
\begin{center}
\epsfig{file=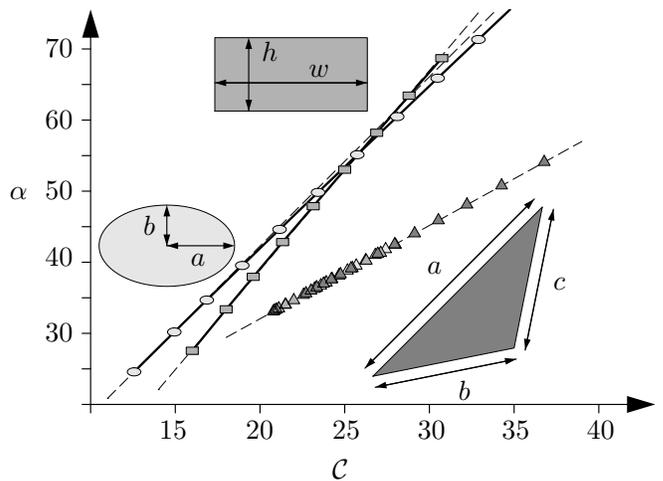, width=\columnwidth,clip}
\end{center}
\caption{Correction factor versus compactness for the elliptical,
rectangular, and triangular classes. The solid lines are the exact
results, and the dashed lines indicate
Eqs.~(\ref{eq:alpha-C-elliptical}),
(\ref{eq:alpha-C-rectangular-approximate}), and
(\ref{eq:alpha-triangle}). Numerical results from a finite-element
simulation are also included ($\circ$, $\square$, and
$\triangle$). Note that in the case of triangles all classes
(right, isosceles, and acute/obtuse scalene triangles --- marked
by different grayscale triangles) fall on the same straight line.}
\label{fig:alpha_vs_C}
\end{figure}

The elliptical family of cross-sections is special in the sense
that Eq.~(\ref{eq:Poisson}) can solved analytically (see e.g.
Ref.~\onlinecite{Landau:87a}) and we can get an explicit
expression for the geometrical correction factor introduced in
Eq.~(\ref{eq:alphaDef}). For an ellipse centered at the origin
with semi-major and minor axes $a$ and $b$ it can be verified by
direct insertion that
 \begin{equation}
 v(x,y)=\frac{\Delta p}{\eta L}\frac{(ab)^2}{2(a^2+b^2)}
 \left(1-\frac{x^2}{a^2}-\frac{y^2}{b^2}\right)
 \end{equation}
fulfils Eq.~(\ref{eq:Poisson}). From Eq.~(\ref{eq:Deltap}) it can
now be shown that
 \begin{equation}\label{eq:Deltap_elliptical}
 \alpha(\gamma)=4\pi(\gamma+\gamma^{-1})
 \end{equation}
where $\gamma=a/b$. Furthermore, for an ellipse we have
 \begin{equation}
 \label{eq:c_elliptical}
 \calC(\gamma)= \frac{16}{\pi}\:
 \gamma\left(\int_0^{\pi/2}d\theta\,
 \sqrt{1-(1-\gamma^{-2})\sin^2\theta}\:\right)^2.
 \end{equation}
The relation between $\alpha$ and $\calC$ can now be investigated
through a parametric plot. In order to get an approximate
expression for $\alpha(\calC)$ we begin by inverting
Eq.~(\ref{eq:Deltap_elliptical}). By selecting the proper root we
get $\gamma(\alpha)$ which we then substitute into
Eq.~(\ref{eq:c_elliptical}) such that
 \begin{equation}\label{eq:c_elliptical_alpha}
 \calC(\alpha)= \frac{1}{2\pi^2}\left(\int_0^\pi d\theta
 \sqrt{\alpha+\sqrt{\alpha^2-(8\pi)^2}\cos\theta}\:\right)^2.
 \end{equation}
Expanding around $\alpha=8\pi$ and inverting we get
 \begin{equation}\label{eq:alpha-C-elliptical}
 \alpha(\calC) = \frac{8}{3}\:\calC-\frac{8\pi}{3}
 +{\cal O}([\calC-4\pi]^2),
 \end{equation}
and in Fig.~\ref{fig:alpha_vs_C} we compare the exact solution
(solid line), from a parametric plot of
Eqs.~(\ref{eq:Deltap_elliptical}) and (\ref{eq:c_elliptical}), to
the approximate result (dashed line) in
Eq.~(\ref{eq:alpha-C-elliptical}). Results of a numerical
finite-element solution of Eq.~(\ref{eq:Poisson}) are also
included ($\circ$ points). As seen, there is a close-to-linear
dependence of $\alpha$ on $\calC$ as described by
Eq.~(\ref{eq:alpha-C-elliptical}).

\subsection{Rectangular cross section}
For a rectangle with width-to-height ratio $\gamma=w/h$ we solve
Eq.~(\ref{eq:Poisson}) using Fourier series \cite{white:03},
 \begin{align}
 v(x,y)&=\frac{\Delta p}{\eta L}
 \frac{4h^2}{\pi^3}\\
 &\times \sum_{n=1,3,5,\ldots}^\infty
 \frac{1}{n^3}\left(1-\frac{\cosh(n\pi x/h)}{\cosh(n\pi
 w/2h)}\right)\sin(n\pi y/h)\nonumber
 \end{align}
is indeed a solution. Here, the coordinate system is chosen so
that $-w/2<x<w/2$ and $0<y<h$. From Eq.~(\ref{eq:Deltap}) it
follows that
 \begin{equation}\label{eq:Deltap-gamma-rectangular}
 \alpha(\gamma)=\frac{\pi^3 \gamma^2}{8}
 \left(\sum_{n=1,3,5,\ldots}^{\infty} \frac{n\gamma}{\pi
 n^5}-\frac{2}{\pi^2 n^5}\tanh(n \pi \gamma/2) \right)^{-1}
 \end{equation}
and for the compactness we have
 \begin{equation}\label{eq:C-gamma-rectangular}
 \calC(\gamma)=8+4 \gamma +4/\gamma.
 \end{equation}
Using that $\tanh(x)\simeq 1$ for $x\gg 1$ we get
 \begin{equation}\label{eq:Deltapapprox}
 \alpha(\gamma) \simeq \frac{12\pi^5 \gamma^2}{\pi^5\gamma-186
 \zeta(5)},\quad\gamma\gg 1,
 \end{equation}
and substituting $\gamma(\calC)$ into this expression and
expanding around $\calC(\gamma=2)=18$ we get
 \begin{equation}\label{eq:alpha-C-rectangular-approximate}
 \alpha(\calC)\approx \frac{22}{7}\:\calC-\frac{65}{3} +
 {\cal O}\big([\calC-18]^2\big).
 \end{equation}
For the two Taylor coefficients we have used the first three terms
in the continued fraction. In Fig.~\ref{fig:alpha_vs_C} we compare
the exact solution, obtained by a parametric plot of
Eqs.~(\ref{eq:Deltap-gamma-rectangular}) and
(\ref{eq:C-gamma-rectangular}), to the approximate result,
Eq.~(\ref{eq:alpha-C-rectangular-approximate}). Results of a
numerical finite-element solution of Eq.~(\ref{eq:Poisson}) are
also included ($\square$ points). As in the elliptical case, there
is a close-to-linear dependence of $\alpha$ on $\calC$ as
described by Eq.~(\ref{eq:alpha-C-rectangular-approximate}).

\subsection{Triangular shape}

For the equilateral triangle it can be shown analytically that
$\alpha=20\sqrt{3}$ and $\calC=12\sqrt{3}$, see e.g.
Ref.~\onlinecite{Landau:87a}. However, in the general case of a
triangle with side lengths $a$, $b$, and $c$ we are referred to
numerical solutions of Eq.~(\ref{eq:Poisson}). In
Fig.~\ref{fig:alpha_vs_C} we show numerical results ($\triangle$
points), from finite-element simulations, for scaling of right
triangles, isosceles triangles, and acute/obtuse scalene triangles
(for the definitions we refer to Ref.~\onlinecite{Weisstein:04a}).
The dashed line shows
 \begin{equation}\label{eq:alpha-triangle}
 \alpha(\calC) = \frac{25}{17}\:\calC+\frac{40\sqrt{3}}{17},
 \end{equation}
where the slope is obtained from a numerical fit and subsequent
use of the first three terms in the continued fraction of this
value. As seen, the results for different classes of triangles
fall onto the same straight line. Since we have
 \begin{equation}
 \calC(a,b,c)=\frac{8(a+b+c)^2}{
 \sqrt{\frac{1}{2}\big(a^2+b^2+c^2\big)^2
 -\big(a^4+b^4+c^4\big)}}
 \end{equation}
the result in Eq.~(\ref{eq:alpha-triangle}) allows for an easy
evaluation of $R_\textrm{hyd}$ for triangular channels.
\begin{figure}[t!]
\begin{center}
\epsfig{file=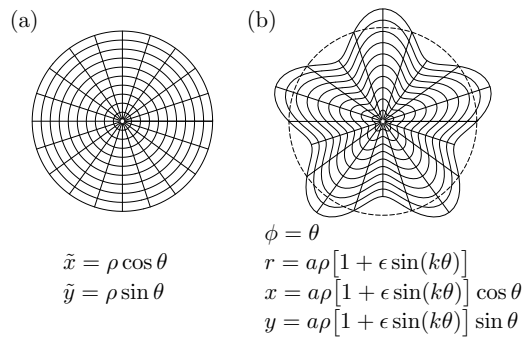, width=0.8\columnwidth,clip}
\end{center}
\caption{(a) The geometry of the unperturbed and analytically
solvable cross section, the unit circle, described by coordinates
$(\tilde{x},\tilde{y})$ or $(\rho,\theta)$. (b) The geometry of
the perturbed cross section described by coordinates $(x,y)$ or
$(r,\phi)$ and the perturbation parameter $\epsilon$. Here $a=1$,
$k=5$ and $\epsilon=0.2$. } \label{fig:multipole}
\end{figure}

\subsection{Harmonically perturbed circle}
By use of shape perturbation theory it is possible to extend the
analytical results for Poiseuille flow beyond the few cases of
regular geometries that we have treated above. In shape
perturbation theory the starting point is an analytically solvable
case, which then is deformed slightly characterized by some small
perturbation parameter $\epsilon$. As illustrated in Fig.
\ref{fig:multipole} the unperturbed shape is described by
parametric coordinates $(\tilde{x},\tilde{y})$ in Cartesian form
or $(\rho,\theta)$ in polar form. The coordinates of the physical
problem we would like to solve are $(x,y)$ in Cartesian form and
$(r,\phi)$ in polar form.

As a concrete example we take the harmonic perturbation of the
circle defined by the transformation
\begin{subequations}
\begin{align}
 \phi &= \theta,\label{eq:Ta}\\
 r  &= a\:\rho\big[1 + \epsilon \sin(k\theta)\big],\label{eq:Tb}\\
 x(\rho,\theta)  &= a\:\rho\big[1 + \epsilon \sin(k\theta)\big]\cos\theta,\\
 y(\rho,\theta)  &= a\:\rho\big[1 + \epsilon \sin(k\theta)\big]\sin\theta,
 \end{align}
\end{subequations}
where $a$ is length scale, $k$ is an integer ($>2)$ defining the
order of the harmonic perturbation, $0\leq \theta\leq 2\pi$, and
$0\leq \rho\leq 1$. For $\epsilon=0$ the shape is unperturbed. The
boundary of the perturbed shape is described by fixing the
unperturbed coordinate $\rho=1$ and sweeping in $\theta$,
 \begin{equation}
 \partial\Omega: \quad \big(x,y\big) =
 \big(x[1,\theta],\: y[1,\theta] \big).
 \end{equation}
It is desirable to formulate the perturbed Poiseuille problem
using the unperturbed coordinates. To obtain analytical results it
is important to make the appearance of the perturbation parameter
explicit. When performing a perturbation calculation to order $m$
all terms containing $\epsilon^l$ with $l>m$ are discarded, while
the remaining terms containing the same power of $\epsilon$ are
grouped together, and the equations are solved power by power. To
carry out the calculation the velocity $v(x,y)$ is written as
 \begin{align}\label{eq:v-expansion}
 v(x,y) &=
 v\big(x[\rho,\theta],y[\rho,\theta]\big)\\
 & = v^{{(0)}}(\rho,\theta) + \epsilon\: v^{{(1)}}(\rho,\theta)
 + \epsilon^2\:  v^{{(2)}}(\rho,\theta) + \cdots\nonumber
 \end{align}
Likewise, the Laplacian operator in Eq.~(\ref{eq:Poisson}) must be
expressed in terms of $\rho$, $\theta$, and $\epsilon$. The
starting point of this transformation is the transformation of the
gradients
\begin{subequations}
 \begin{align}
 \partial_r &= (\partial_r \rho)\:\partial_\rho +
 (\partial_r \theta)\:\partial_\theta,\label{PertGradR}\\
 \partial_\phi &= (\partial_\phi \rho)\:\partial_\rho
 + (\partial_\phi \theta)\:\partial_\theta.\label{PertGradPhi}
 \end{align}
 \end{subequations}
The derivatives $(\pp_r \rho)$, $(\pp_r \theta)$, $(\pp_\phi
\rho)$, and $(\pp_\phi \theta)$ are obtained from the inverse
transformation of Eqs.~(\ref{eq:Ta}) and (\ref{eq:Tb}). The
expansion in Eq.~(\ref{eq:v-expansion}) can now be inserted into
Eq.~(\ref{eq:Poisson}) and using the derivatives,
Eqs.~(\ref{PertGradR}) and (\ref{PertGradPhi}), we can carry out
the perturbation scheme. The calculation of the velocity field to
fourth order is straightforward, but tedious. With the velocity
field at hand we can calculate the flow rate and from
Eq.~(\ref{eq:Deltap}) we get
  \begin{align}\label{eq:alpha_epsilon}
 \alpha&=8\pi\Bigg[1+2(k-1)\:\epsilon^2 \\
 &\quad\quad\quad+\frac{47-78k+36k^2-4k^3}{8}\:\epsilon^4\Bigg]+
 {\cal O}\big(\epsilon^6\big),\nonumber
 \end{align}
where we have used the exact result $\calA = \big(1 +
\tfrac{1}{2}\:\epsilon^2\big)\:\pi a^2$ for the area. The result
only involves even powers of $\epsilon$ since
$\epsilon\rightarrow-\epsilon$ is equivalent to a shape-rotation,
which should leave $\alpha$ invariant. From an exact calculation
of the perimeter $\calP$ we get the following expression for
$\calC$,
 \begin{equation}
 \calC = 4\pi +
 2\pi(k^2-1)\:\epsilon^2.
 \end{equation}
Since $\alpha$ is also quadratic in $\epsilon$ this means that
$\alpha$ depends linearly on $\calC$ to fourth order in
$\epsilon$,
 \begin{equation}
 \alpha(\calC)= \frac{8}{1+k}\:\calC - 8\frac{3-k}{1+k}\:\pi
 + {\cal O}\big(\epsilon^4\big).
 \end{equation}
Note that although derived for $k>2$ this expression coincides
with that of the ellipse, Eq.~\eqref{eq:alpha-C-elliptical}, for
$k=2$. Comparing Eq.~(\ref{eq:alpha_epsilon}) [to second order in
$\epsilon$] with exact numerics we find that for $\epsilon$ up to
0.4 the relative error is less than 0.2\% and 0.5\% for $k=2$ and
$k=3$, respectively.

\section{Discussion and conclusion}
\label{sec:conclusion}

We have considered pressure-driven, steady state Poiseuille flow
in straight channels with various shapes, and found a
close-to-linear relation between $\alpha$ and $\calC$. Since the
hydraulic resistance is $\Rhyd \equiv \alpha \RhydS$, we conclude
that $\Rhyd$ depends linearly on $\calC\RhydS$. Different classes
of shape all display this linear relation, but the coefficients
are non-universal. However, for each class only two points need to
be calculated to fully specify the relation for the entire class.
The difference is due to the smoothness of the boundaries. The
elliptical and harmonic-perturbed classes have boundaries without
any cusps whereas the rectangular and triangular classes have
sharp corners. The over-all velocity profile tends to be convex
and maximal near the center-of-mass of the channel, see
Fig.~\ref{fig:geometry}. If the boundary is smooth the velocity in
general goes to zero in a convex parabolic manner whereas a
concave parabolic dependence is generally found if the boundary
has a sharp corner (this can be proved explicitly for the
equilateral triangle \cite{Landau:87a}). Since the concave drop is
associated with a region of low velocity compared to the convex
drop, geometries with sharp changes in the boundary tend to have a
higher hydraulic resistance compared to smooth geometries with
equivalent cross-sectional area.

We believe that the explicit and simple link between $R_{\rm hyd}$
and $\calC$ is an important observation since at the same time
$\calC$ is also central to the strength and effectiveness of
various surface-related phenomena. We note that in micro-channels
the flow properties and electrokinetic phenomena may be somewhat
connected and substantial deviations from classical Poiseuille
flow have been reported recently, see Ref.~\onlinecite{Phares:04a}
and references therein. Nevertheless, our observation is an
important first step with relevance to the use of micro-fluidic
channels in lab-on-a-chip applications. Furthermore, our results
allow for an easy evaluation of the hydraulic resistance for
elliptical, rectangular, and triangular cross-sections with the
geometrical measure $\calC$ being the only input parameter.

\section*{Acknowledgement}
We thank J. Kutter for stimulating discussions.  N.~A.~M. and
F.~O. are supported by The Danish Technical Research Council
(Grants No.~26-03-0073 and No.~26-03-0037).


\expandafter\ifx\csname
natexlab\endcsname\relax\def\natexlab#1{#1}\fi
\expandafter\ifx\csname bibnamefont\endcsname\relax
  \def\bibnamefont#1{#1}\fi
\expandafter\ifx\csname bibfnamefont\endcsname\relax
  \def\bibfnamefont#1{#1}\fi
\expandafter\ifx\csname citenamefont\endcsname\relax
  \def\citenamefont#1{#1}\fi
\expandafter\ifx\csname url\endcsname\relax
  \def\url#1{\texttt{#1}}\fi
\expandafter\ifx\csname
urlprefix\endcsname\relax\def\urlprefix{URL }\fi
\providecommand{\bibinfo}[2]{#2}
\providecommand{\eprint}[2][]{\url{#2}}

\end{document}